\documentclass[8.5pt,twoside,twocolumn]{article}
\usepackage[super,sort&compress,comma]{natbib} 
\usepackage{mhchem}
\usepackage{times,mathptmx}
\usepackage{subfigure}

\usepackage{sectsty}
\usepackage{balance} 

\usepackage{graphicx} 
\usepackage{lastpage}
\usepackage[format=plain,justification=raggedright,singlelinecheck=false,font=small,labelfont=bf,labelsep=space]{caption} 
\usepackage{fancyhdr}
\begin{document}

\thispagestyle{plain}
\fancypagestyle{plain}{
\renewcommand{\headrulewidth}{1pt}}
\renewcommand{\thefootnote}{\fnsymbol{footnote}}
\renewcommand\footnoterule{\vspace*{1pt}%
\hrule width 3.4in height 0.4pt \vspace*{5pt}} 
\setcounter{secnumdepth}{5}

\makeatletter 
\def\subsubsection{\@startsection{subsubsection}{3}{10pt}{-1.25ex plus -1ex minus -.1ex}{0ex plus 0ex}{\normalsize\bf}} 
\def\paragraph{\@startsection{paragraph}{4}{10pt}{-1.25ex plus -1ex minus -.1ex}{0ex plus 0ex}{\normalsize\textit}} 
\renewcommand\@biblabel[1]{#1}            
\renewcommand\@makefntext[1]%
{\noindent\makebox[0pt][r]{\@thefnmark\,}#1}
\makeatother 
\renewcommand{\figurename}{\small{Fig.}~}
\sectionfont{\large}
\subsectionfont{\normalsize} 

\fancyfoot{}
\fancyfoot[RO]{\footnotesize{\sffamily{1--\pageref{LastPage} ~\textbar  \hspace{2pt}\thepage}}}
\fancyfoot[LE]{\footnotesize{\sffamily{\thepage~\textbar\hspace{3.45cm} 1--\pageref{LastPage}}}}
\fancyhead{}
\renewcommand{\headrulewidth}{1pt} 
\renewcommand{\footrulewidth}{1pt}
\setlength{\arrayrulewidth}{1pt}
\setlength{\columnsep}{6.5mm}
\setlength\bibsep{1pt}

\twocolumn[
  \begin{@twocolumnfalse}
\noindent\LARGE{\textbf{Stability of multi-lamellar lipid tubules in excess water}}
\vspace{0.6cm}

\noindent\large{\textbf{Tripta Bhatia}\vspace{0.5cm}}


\vspace{0.6cm}

\noindent \normalsize{In the lyotropic phase of lipids with excess water, multilamellar tubules (MLTs) grow from defects. A phenomenological model for the stability of MLTs is developed that is universal and independent of the underlying growth mechanisms of MLTs. The stability of MLTs implies that they are in hydrostatic equilibrium and stable as elastic objects that have compression and bending elasticity. The results show that even with 0.1 atm solvent pressure differences, the density profile is not significantly altered, thus determining that the stability is due to the trapped solvent. The results are of sufficient value in relation to lamellar stability models and may have implications beyond the described MLT models, especially in other models of membrane systems.}
\vspace{0.5cm}
 \end{@twocolumnfalse}
  ]


\footnotetext{\textit{~Department of Physical Sciences, Indian Institute of Science Education and Research Mohali, Sector 81, Knowledge City, Manauli, SAS Nagar, Punjab, 140306 India. }}

\section*{Introduction}
\begin{figure}[ht!]
\centering
\includegraphics[scale=0.4]{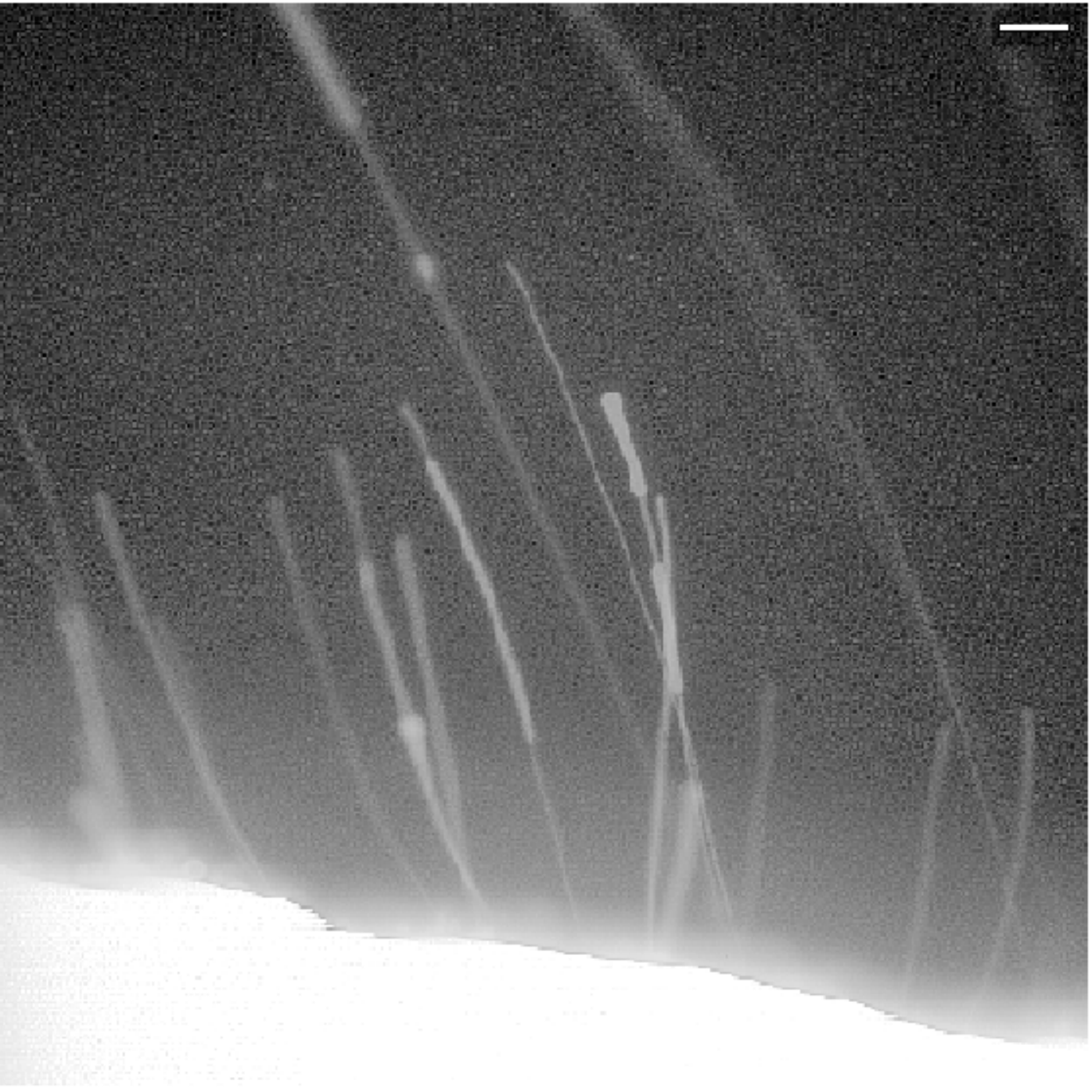}
\caption{\label{figure0} Two-dimensional view of multilamellar tubules (MLTs) dispersed in excess water. The brightest region is the lipid reservoir in which the tubules are rooted. The image was obtained by epi-fluorescence microscopy. The scale bar is $20 \ \mu$m.}

\end{figure}

The curvature of biomembranes and the molecular mechanisms that drive the generation of curvature are important. \cite{LG2016,RM1977,RL12022, TB2022} Bilayer tubules are observed in lipid-protein compartments such as giant unilamellar vesicles (GUV), in vitro. In the GUV system, one of the mechanisms for the generation of curvature is the compositional asymmetry of the leaflet due to the different amounts of glycolipid GM1.\cite{TB20182} Bilayers can also acquire transbilayer asymmetry through asymmetric adsorption (or desorption) layers formed by the composition of the solution, resulting in the generation of membrane shapes in the form of tubules or buds.\cite{TB12020,RL22022,RL12022} This article describes MLTs as quasi-equilibrium structures using elasticity theory.  MLTs are made up of only one lipid 1,2-dioleoyl-sn-glycero-3-phosphocholine (DOPC) and grow from defects. This implies that the bilayers have the same lipid composition and no asymmetry in terms of this composition. Fig.~\ref{figure0} shows an epi-fluorescence image of stable cylindrical MLTs rooted in the lamellar reservoir for which stability conditions are discussed. 

\begin{figure}[hb!]
\centering
\includegraphics[scale=0.7]{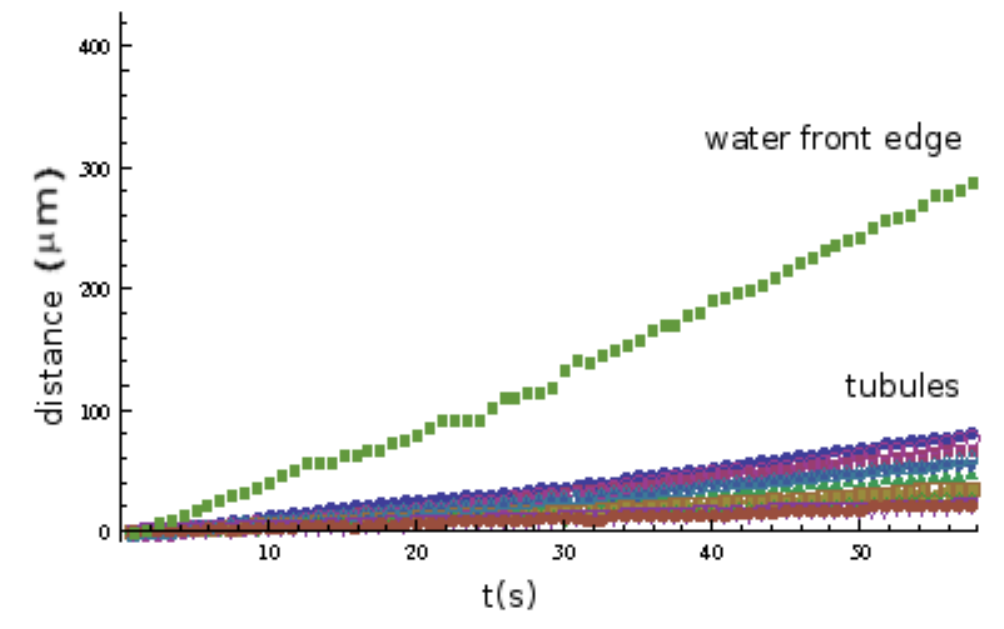}
\centering
\caption{\label{fig.1} Growth of tubules. After the addition of water, the growth of the MLT tip is tracked. The edge of the water front moves at a speed $\sim 5 \ \mu$m/s much faster than the growth speed of MLTs.}
\end{figure}
\section*{Experimental methods}
We have prepared lamellar stacks of DOPC lipid (purchased from Sigma). Approximately 20-50 $ \mu $ l of chloroform lipid solution (containing approximately 0.2 mole\% of membrane dye, 1,2-dihexadecanoyl-sn-glycero-3-phosphoethanolamine, triethylammonium salt, RhPE purchased from Molecular Probes) is spread on a glass coverlip. The sample is gently dried in a nitrogen stream and kept covered inside a desiccator overnight with little vacuum sufficient to hold the chamber tight for the duration. The coated cover slip with dried sample was glued to a larger cover slip at the edges using mica spacers of about 100 $ \mu $m thickness. Solvent was introduced between the coverslips of the sample cells by capillary action. After the dry lamellar stack is hydrated, water molecules are constantly exchanged between the sample and the bulk; therefore, the hydration gradient acts as a driving force for tubule growth, as previously described in detail. \cite{TB2021, BHM} Fig.~\ref{fig.1} shows tubule growth after water is added. The edge of the water front moves at a speed $\sim 5 \ \mu$m/s much faster than the growth of the tip of the MLTs rooted in the reservoir. There are two important results of the previous studies: (i) the need to obtain a necessary first step for the generation of MLTs is the presence of defects; and (ii) once such a membrane tubule has formed, the open chamber needs to be closed to stop all growth and retraction dynamics, thereby achieving hydrostatic equilibrium. \\
The open edges were sealed with silicone glue immediately after the solvent filled the entire gap. We found that in the \textit{sealed} sample cells, an osmotic equilibrium is reached between the swollen lamellar stack and the excess water and the tubes remain stable for a couple of days. Sample cells are observed under a confocal microscope (LEICA TCS-SP2, He-Ne laser 543 nm)  equipped with a 40x dry objective (0.85 N.A.) having a correction collar. Fig.~\ref{fig.2} shows a two-dimensional confocal cross section of the tip and root of the tubules. The double arrow indicates the incident laser polarization. In a closed (sealed) chamber, the diameter of the core of these MLTs is extracted from quantitative image analysis, namely optimum smoothening.\cite{TB2018} Fluorescence confocal polarizing microscopy (FCPM) observations confirmed that the tubules are multilamellar, are capped at the end and originate from defects. The bright region in the image of the tubules is the region where the dye molecules have their absorption and emission dipole moments oriented parallel to the incident laser polarization. For a closed (sealed) chamber, water-molecule exchange can occur only within the sample cell, and an equilibrium is reached for which the number of water molecules leaving the lamellar stack is exactly balanced by the number of water molecules entering the stack, and no further growth of tubules occurs. In this situation, the image reveals that MLTs have a wide range of $r_{\rm{c}}$ and $r_{\rm{o}}$. Given the earlier observations \cite{BHM}, in particular the long lifetime of the structures, it is reasonable to consider these as quasi-equilibrium structures.
\section*{Elasticity of tubules}
\begin{figure}[ht!]
\centering
\begin{subfigure}
\centering
\includegraphics[scale=1.9]{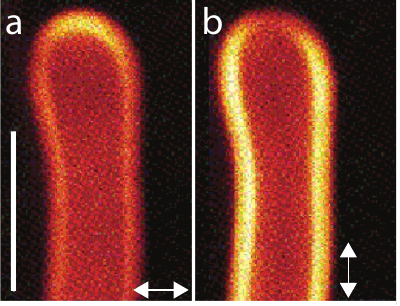}
\end{subfigure} \\
\begin{subfigure}
\centering
\includegraphics[scale=0.42, angle =180]{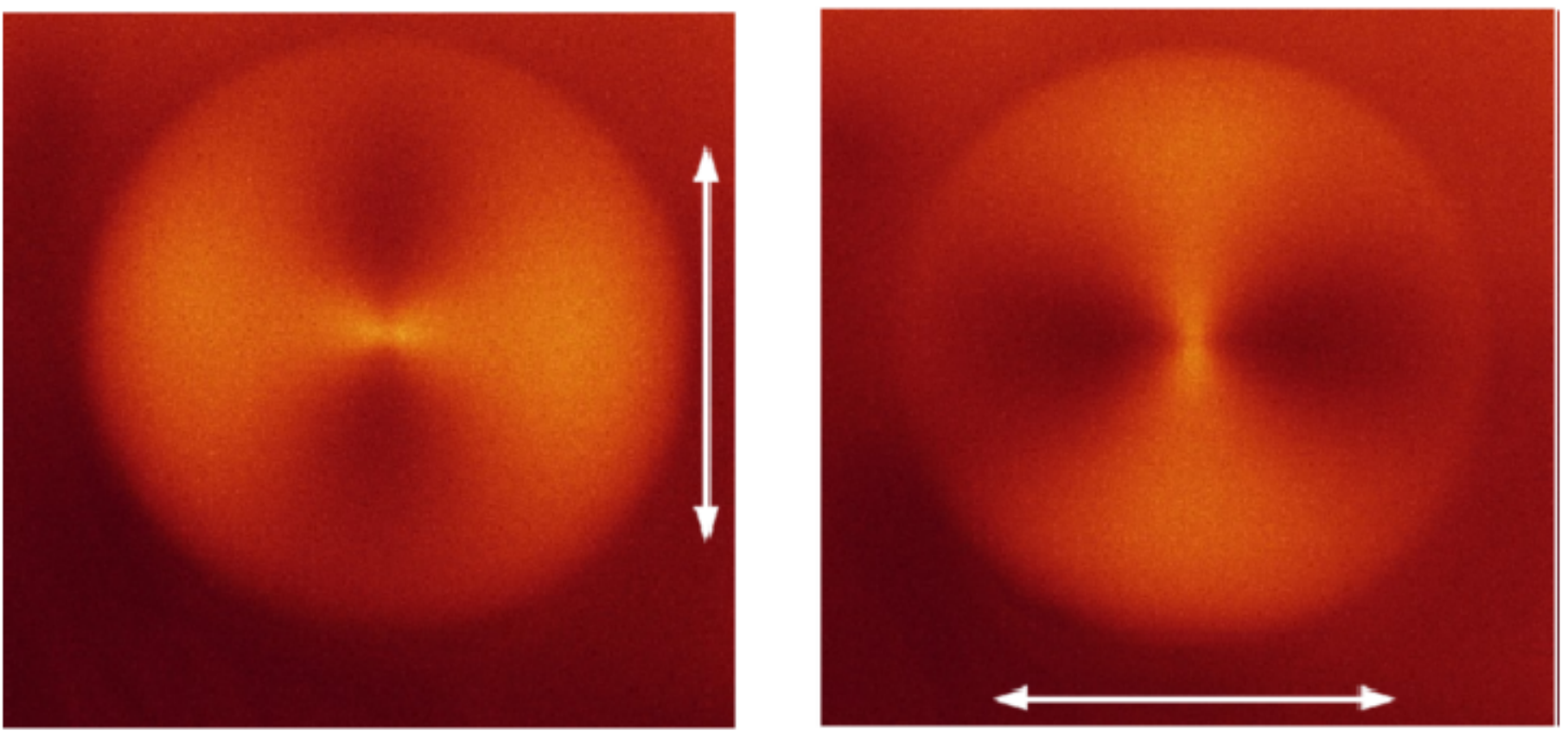}
\end{subfigure}
\caption{\label{fig.2} Two-dimensional confocal cross section of the tip and root of the tubules. The double arrow indicates the incident laser polarization. FCPM observations confirmed that the tubules are multilamellar, are capped at the end, and originate from defects. Top panel: Multilamellar tubule (MLT) observed by fluorescent confocal polarized microscopy (FCPM) with incident laser polarization (a) oriented perpendicular to the long axis of the tubule and (b) parallel to the long axis of the tube. The scale bar is $10 \ \mu$m applied to (a,b). Bottom panel: FCPM image of the root of arbitrary MLTs in the lipid reservoir, which confirms the multilamellar arrangement. The bright region of the tubules is the region where the dye molecules have their absorption and emission dipole moments oriented parallel to the incident laser polarization.}
\end{figure}

A phenomenological model is proposed to analyze the stability of simple cylindrical MLTs with a uniform cross section shown in Fig.~\ref{fig.2}. The pertinent experimental observations on which the model is based are listed below.
\begin{enumerate}
\item After the sample cell is sealed, the tubules remain stable for more than one day.  \cite{BHM,TB20182,TB2021} The model applies to the equilibrium shapes achieved by the MLTs shown in Fig.~\ref{fig.2}.
\item The tubules are capped at the end (Fig.~\ref{fig.2}). The membrane fluorescence signal comes from dye molecules oriented in the bilayer with the absorption and emission dipole moment always oriented in the plane of the bilayer \cite{BHM, TB2021}. Thus, all MLT images exhibit this optical effect, namely photoselection, from which the membrane morphology of MLTs with a well-defined tip can be defined. 
\item The tubules originate from defects in the lipid reservoir, shown in Fig.~\ref{fig.2}. \cite{BHM}
\item Tubules have a wide range of $r_{\mathrm{c}}$ and $r_{\mathrm{o}}$. In general, the examined MLTs have $r_{\rm{o}}$ in the range of 1-30 $\mu$m. Furthermore, in the stability regime, the structures are investigated with a z-stack confocal that detects the diameter of the core in the range $88\pm23$ nm and $6860\pm50$ nm, respectively, using a special image processing technique, namely optimal smoothening. \cite{TB2018}
\end{enumerate}
Given the above observations and, in particular, the long life times of the structures, it is reasonable to consider these as quasi-equilibrium structures. Therefore, the theory of elasticity is used to understand the stability and structure of MLTs.
\paragraph*{Elasticity of stack of bilayers:} The smectic liquid crystals are one-dimensional \textquotedblleft solids\textquotedblright  \hspace{0.5mm} composed of fluid layers exhibiting quasi-long-range order in the orthogonal direction of the layers. For small distortions, the elastic free energy (in Cartesian coordinates) is \cite{deGP,CL,WH1973}.
\begin{equation}
F_{\mathrm{el}}\hspace{0.5mm}= \int \hspace{0.5mm}\Big [ \frac{B}{2} \hspace{0.5mm}\Big ( \partial_{z} u \Big )^{2}\hspace{0.5mm} + \hspace{0.5mm}\frac{\kappa}{2}\hspace{0.5mm} H^{2}\hspace{0.5mm}+ \hspace{0.5mm}\kappa_{\mathrm{G}}\hspace{0.5mm} K  \Big ]\hspace{0.5mm} dx \hspace{0.5mm}dy\hspace{0.5mm} dz \hspace{0.5mm} , \label{elastsm}
\end{equation}
where $ z $ is the solid direction \textquotedblleft like\textquotedblright \hspace{0.5mm} of the layering, $u\hspace{0.5mm}(x,y,z)$ is the layer displacement field. $(B/2) \hspace{0.5mm}\Big ( \partial_{z} u \Big )^{2}$ is the energy density for compression (or extension) of the layer spacing with $B$ as the compression modulus. $(\kappa/2)\hspace{0.5mm} H^{2}$ is the energy density for layer bending, with the modulus of bending $\kappa$ and the mean curvature $H=[(1/2)(1/R_{1}\hspace{0.5mm} +\hspace{0.5mm}1/R_{2})=\bigtriangledown_{\perp} ^{2} u]$, with the principal radii of curvature $R_{1}$ and $R_{2}$ and $\bigtriangledown_{\perp} ^{2}=(\partial_{x}^{2} + \partial_{y}^{2})$. $K=(1/R_{1}R_{2})$ is the Gaussian curvature. The Gaussian curvature term contributes to the energy only if the system under study has a boundary or undergoes a change in topology (number of handles). 
The tubule in Fig.~\ref{fig.2} has a single core, with a uniform core radius $(r_{\rm{c}})$ and an outer radius $(r_{\rm{o}})$. The edge structure of the tubule is probed by changing the incident laser polarization by $(\pi /2)$ from Fig.~\ref{fig.2}a to Fig.~\ref{fig.2}b, which confirms that the tubule is capped at the end and that the open edges of the bilayers are not exposed to the solvent. MLTs grow from defects in the hydrated lamellar stack \cite{BHM}. The main result of our paper is to develop a curvature-elastic model for the stability of MLTs, as described in the following. 
\paragraph*{Elasticity of MLTs:} An MLT is an aggregate of multiple bilayers that are rolled coaxially to form cylindrical structures characterized by an outer radius $r_{\rm{o}}$ and an inner core radius $r_{\rm{c}}$ that are dispersed in the bulk water in the sample cell. The tubules are capped at one end, and the other end is rooted into the lamellar stack. We have ignored the tubule cap energy, since the tubules are $(10-100) \ \mu$m long. The elastic energy of the tubular lamella contributes from the compression of the layer and the curvature of the layer \cite{deGP,CL} with the direction of the layer changed from $z \rightarrow r$ and the direction of the in-plane changed from $(x,y)\rightarrow(r,\phi)$, where $r$ is the radial direction, $\phi$ is the polar angle and $z$ lies parallel to the long axis of the tubules. Equation (\ref{elastsm}) becomes

\begin{equation}
F_{\mathrm{el}}\hspace{0.5mm}= \int \hspace{0.5mm}\Big [ \frac{B}{2} \hspace{0.5mm}\Big ( \partial_{r} u \Big )^{2}\hspace{0.5mm} + \hspace{0.5mm}\frac{\kappa}{2}\hspace{0.5mm} \left(\partial^2 _{r} u + \frac{1}{r^2}\partial^2 _{\phi} u + \frac{1}{r}\partial_{r} u \right)^2\hspace{0.5mm} \Big ]\hspace{0.5mm} r dr \hspace{0.5mm}d\phi\hspace{0.5mm} dz \hspace{0.5mm} 
\end{equation}

where the symbols have their usual meaning, $\partial_{r}$ stands for the differential operator d/d$r$ and $H= \left(\partial^2 _{r} u + \frac{1}{r^2}\partial^2 _{\phi} u + \frac{1}{r}\partial_{r} u \right)$ is the mean curvature. The Gaussian curvature term is neglected \cite{SS}. We retain the subdominant term in the free elastic energy because the compressive and curvature stresses must be balanced for stability. We define $\lambda^{2}=(\kappa / B)$ where $\lambda$ and $B$ have the dimension of length and pressure, respectively.

If the tube is assumed to have a uniform circular cross section along its length, then the dependence of $u(r,\phi,z)$, on $\phi$ becomes zero for hydrostatic equilibrium, i.e; $(\partial_{\phi} u =0)$. Hence, for cylindrically symmetric MLTs with a uniform cross section, the elastic free energy is;
\begin{equation*}
\begin{split}
{F}_{\rm{el}} & = \int_{r,z,\phi} \ r \frac{B}{2} \left[ (\partial_{r} u)^2 + \lambda^2 \left(\partial^2 _{r} u + \frac{1}{r}\partial_{r} u \right)^2 \right] d r \ d \phi \ d z \\ 
&= 2 \pi \int_{r,z} \ r f[u(r), u'(r),u''(r)] d r \ d z
\end{split}
\end{equation*}
where $u'(r)=\partial_{r} u, \ u''(r)=\partial^2 _{r} u$ and $F_{\rm{el}}$ have the dimension of energy. We consider all possible variations in ${F}_{\rm{el}}$ induced by a shift in layer displacement from $u$ to $(u+\varDelta u)$,
\begin{equation}
\begin{split}
\delta  {F}_{\rm{el}} &  = 2 \pi \int_{r,z} \delta \left(r f [u(r), u'(r),u''(r)] \right)  d r \ d z \\
& = 2 \pi \int_{r,z} \left[\frac{\partial (r f)}{\partial u} \delta u +  \frac{\partial (r f)}{\partial u'(r)} \delta u'(r) +  \cdot \cdot \cdot \right]  d r \ d z 
\end{split}
\end{equation}
 
\subsection*{Condition for stability}
\begin{figure}[ht]
\centering
\includegraphics[scale=0.55]{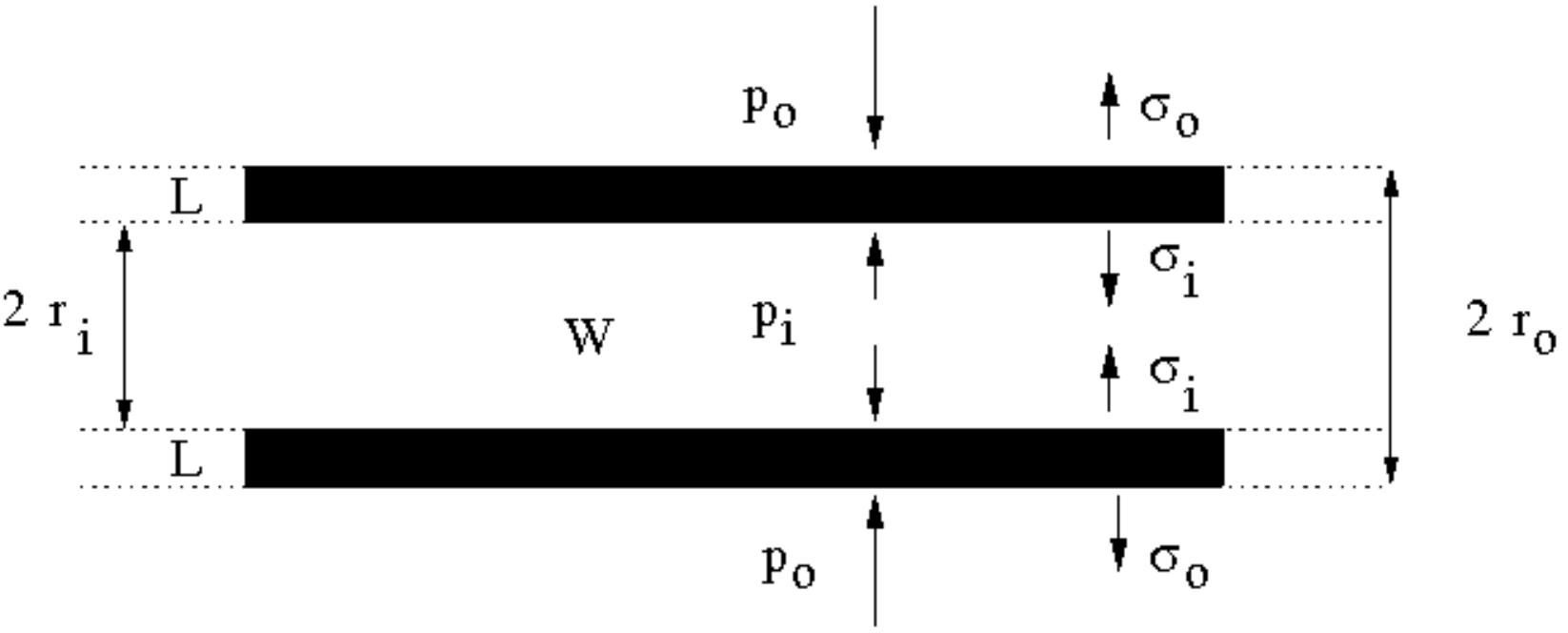}
\caption{\label{fig.3} An MLT in hydrostatic equilibrium. $\sigma_{\rm{i}}$ and $\sigma_{\rm{o}}$ are the normal stresses in the innermost and outermost layer that counterbalance the pressures $\rm{p}_{\rm{i}}$ and $\rm{p}_{\rm{o}}$ of the solvent. "W" denotes the water (or solvent), "L" denotes the lamella, $r_{\rm{i}}(=r_{\rm{c}})$ and $r_{\rm{o}}$ denote the inner (or core) and outer radii of the tubule, respectively.} 
\end{figure}

The tubules are bounded by water at the top, between the bilayers, and inside the core. To be in hydrostatic equilibrium, the normal stresses at $r_{\rm{c}}$ and $r_{\rm{o}}$ must balance the fluid pressure inside and outside the tubule, respectively, as shown in Fig.~\ref{fig.3}. If the negative and positive signs show the direction outward and inward from the bilayer plane, respectively, then 
\begin{equation} 
\label{bc}
\begin{split}
\sigma (r_{\rm{i}}) = - p_{\rm{i}},  \\
- \ \sigma (r_{\rm{o}}) = p_{\rm{o}} \
\end{split}
\end{equation}
where $\sigma_{\rm{i}}$ and $\sigma_{\rm{o}}$ are the normal stresses in the innermost and outermost layers that counterbalance the pressures $p_{\rm{i}}$ and $p_{\rm{o}}$ of the solvent in $r_{\rm{i}}$ and $r_{\rm{o}}$, respectively, as shown in Fig.~\ref{fig.3}.  
In equilibrium, the Euler-Lagrange equation $(\delta F_{\rm{el}} / \delta u)=0$ holds within the bulk of the lamellar region of the MLT giving, 
\begin{equation}
\begin{split}
\left(\lambda ^2-r^2\right) u'(r) 
&-r \left(\lambda^2+r^2\right) u''(r) \\
&+ r^2 \lambda^2 \left(r u^{(4)}(r)+2 u^{(3)}(r) \right)=0
\end{split}
\end{equation}

The solution to the Euler-Lagrange equation for $(r>0, \lambda>0)$ is given by
\begin{equation}
\begin{split}
u(r) &= c_3 \ \lambda ^2 Y_0\left(-\frac{i r}{\lambda }\right)+c_2 \ \lambda ^2
   \left(I_0\left(\frac{r}{\lambda }\right)-1\right)+c_1 \ \log (r)+c_4 \nonumber \\
&= c_2 \ \lambda ^2 \left(I_0\left(\frac{r}{\lambda }\right)-1\right)+c_1 \log (r)+c_4,    
\end{split}
\end{equation}  

where $I_{n}(x)$ and $Y_{\rm{n}}(x)$ \cite{AS}, respectively, represent modified Bessel functions of the first kind and Bessel functions of the second kind and of order $n$, and the four integration constants are denoted with the symbol $ c $. The layer displacement $u (r)$ is real. Therefore, we drop the term proportional to $ Y_{0}\left(-i r / \lambda \right) $, which is a pure imaginary. The constant $ c_{4} $ corresponds to the uniform and rigid displacements of the entire MLT and can therefore be set to zero $ (c_4=0) $ by an appropriate choice of the origin of the coordinate system. This gives a certain real value of $u (r)$.
\begin{equation}
\label{ue}
u(r)\simeq c_2 \ \lambda ^2 \left(I_0\left(\frac{r}{\lambda }\right)-1\right)+c_1 \log (r)
\end{equation}
The gradient of layer displacement or the bilayer compression, $\partial_{r} u(r)$ is a direct measure of the density change given by  
\begin{equation}
v (r) = \partial_{r} u(r)=c_2 \ \lambda \ I_1\left(\frac{r}{\lambda }\right)+\frac{c_1}{r}
\end{equation}
Now, it remains to find the two integration constants $ c_{1} $ and $ c_{2} $ subject to the boundary conditions given by equation (\ref{bc}).  

\subsection*{Solving for $u(r)$}
The divergence of normal stress \cite{LLPK} $(\partial_{i} \sigma_{ij})$ on a patch of membrane with area ($dA =r dr d\phi$) and volume ($dV =r dr d\phi d z$) is related to the total force per unit volume given by;

\begin{equation}
f_{tot}=\int_{V} ( \partial_{i} \sigma_{ij} ) \ d V=\int_{A} \sigma_{ij} \ \hat{n}_{j} \cdot d \hat{A}
\end{equation}

where $\hat{n}_{j}$ is the normal area vector. Normal stress $\sigma(r)$ and radial strain $u'(r)$ are related to elastic free energy \cite{LLPK} as;
\begin{equation} 
\label{stres}
\begin{split}
\sigma (r) &= \frac{\delta {F}_{\rm{el}}}{\delta(u'(r))} = \frac{1}{r} \frac{\delta (r f)}{\delta(u'(r))}\\
&= \frac{2 \pi  B}{r}  \left(\lambda ^2 u''(r)+\left(\frac{\lambda ^2}{r}+r\right)
   u'(r)\right) \\
&= \frac{2 \pi  B}{r}  \left(\lambda ^2 v'(r)+\left(\frac{\lambda ^2}{r}+r\right)
   v(r)\right)    
\end{split}
\end{equation}

Solving equations (\ref{bc}) and (\ref{stres}) we obtain the following. 
\begin{equation}
\begin{split}
c_{1} &= r_{\rm{i}} \left[\frac{p_{\rm{i}} r_{\rm{i}}^2}{2 \pi  B \alpha}+c_{2} \lambda  \left(\frac{\lambda  r_{\rm {i}}
   I_{0}\left(\gamma_{\rm{i}} \right)}{\alpha}-I_{1}\left(\gamma_{\rm{i}} \right)\right)\right], \nonumber \\
c_{2} &= \frac{p_{\rm{i}} r_{\rm{i}}^3 \beta -p_{\rm{o}} r_{\rm{o}}^3 \alpha}{2 \pi  B \lambda  \left[\alpha \left(\beta
   \left(r_{\rm{i}} I_{1} \left(\gamma_{\rm{i}} \right) -r_{\rm{o}} I_{1} \left(\gamma_{\rm{o}} \right) \right)+\lambda  r_{\rm{o}}^2
   I_{0} \left(\gamma_{\rm{o}} \right)\right)-\lambda 
   r_{\rm{i}}^2 \beta
   I_{0} \left(\gamma_{\rm{i}} \right)\right]} \nonumber
\end{split}
\end{equation}

where $\alpha=\left(2 \lambda^2+r_{\rm{i}}^2\right)$, $\beta=\left(2 \lambda^2+r_{\rm{o}}^2\right)$, $\gamma_{\rm{i}}=\left( r_{\rm{i}}/ \lambda \right)$, and $\gamma_{\rm{o}}=\left( r_{\rm{o}}/ \lambda \right)$. Substituting $ c_{1} $ and $ c_{2} $ into equation (\ref{ue}), $u(r)$ is calculated from equation (\ref{ue}).

\subsection*{Hydrostatic equilibrium}
\begin{figure}[ht]
\centering
\includegraphics[scale=0.7]{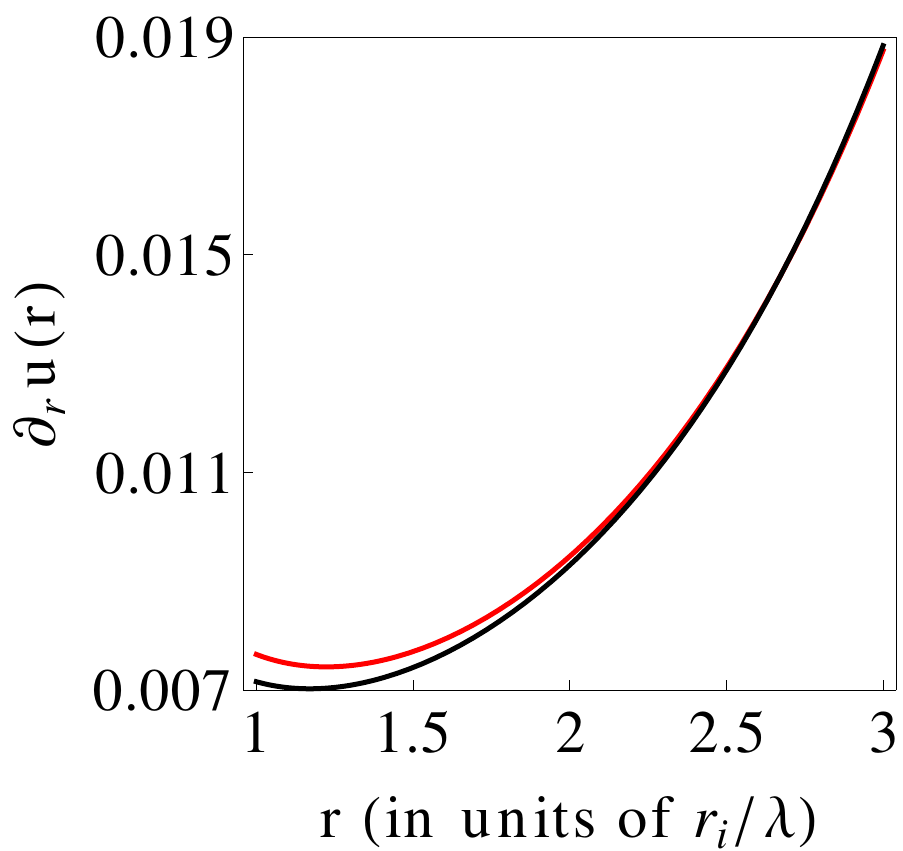}
\caption{\label{fig.4}Red curve [$(\rm{p}_{\rm{i}}-\rm{p}_{\rm{o}})=0.1\ $ atm], black curve ($\rm{p}_{\rm{i}}=\rm{p}_{\rm{o}}$). The tubule has $(r_{\rm{c}},\ r_{\rm{o}})=(1, 3) $ in dimensionless units of length as $(r_{\rm{i}}/\lambda)$.} 
\end{figure}

The bilayer compression is plotted for an MLT with a given pressure difference $(p_{\rm{i}}-p_{\rm{o}})$ of the solvent. I have chosen the dimensionless unit of length as $\gamma_{\rm{i}}= (r_{\rm{i}}/\lambda) = 1 $, and measure the pressure in units of $ B $. For the $ L_{\alpha} $ phase under consideration, $ \lambda $ is of the order of a few layer spacings and $ B \simeq \ 10 $ atm.
\\ The $\partial_{r} u(r)$ for an MLT with the same inner and outer pressures $( p_{\rm{i}}=p_{\rm{o}})$ of the solvent is plotted and shows that the MLT has varying non-zero bilayer compression, as shown in Fig.~\ref{fig.4} by the black color curve. If the inner and outer pressures are zero, the bilayer compression, $\partial_{r} u(r)$ and the stresses are zero throughout the tubule. 

 In reality, the solvent pressures inside and outside the tubule can be different, because $(r_{\rm{c}}<r_{\rm{o}})$. A lower $r_{\rm{c}}$ implies a higher normal stress of the layer curvature (the mean curvature is 1/2$r_{\rm{c}}$ at the core-solvent interface). The normal stress at the outer radius is clearly lower, as $(r_{\rm{o}}>r_{\rm{c}})$. Thus, it is reasonable to assume that the solvent pressure inside the tubule (in the core) is greater than that outside the tubule. We have plotted $\partial_{r} u(r)$ for an MLT with unequal inner and outer pressures $( p_{\rm{i}} \neq p_{\rm{o}})$ of the solvent, e.g., $(p_{\rm{i}}-p_{\rm{o}})=0.1\ $ atm. Even in this extreme case, we find that the compression and, therefore, the density profile of the lamellar material within the MLT do not change significantly, as shown in Fig.~\ref{fig.4}, by the red color curve. In such a scenario, the pressure difference can be stabilized by trapping the solvent within small closed regions that are dynamically formed in the reservoir.

\begin{figure}[ht!]
\centering
\includegraphics[scale=0.4]{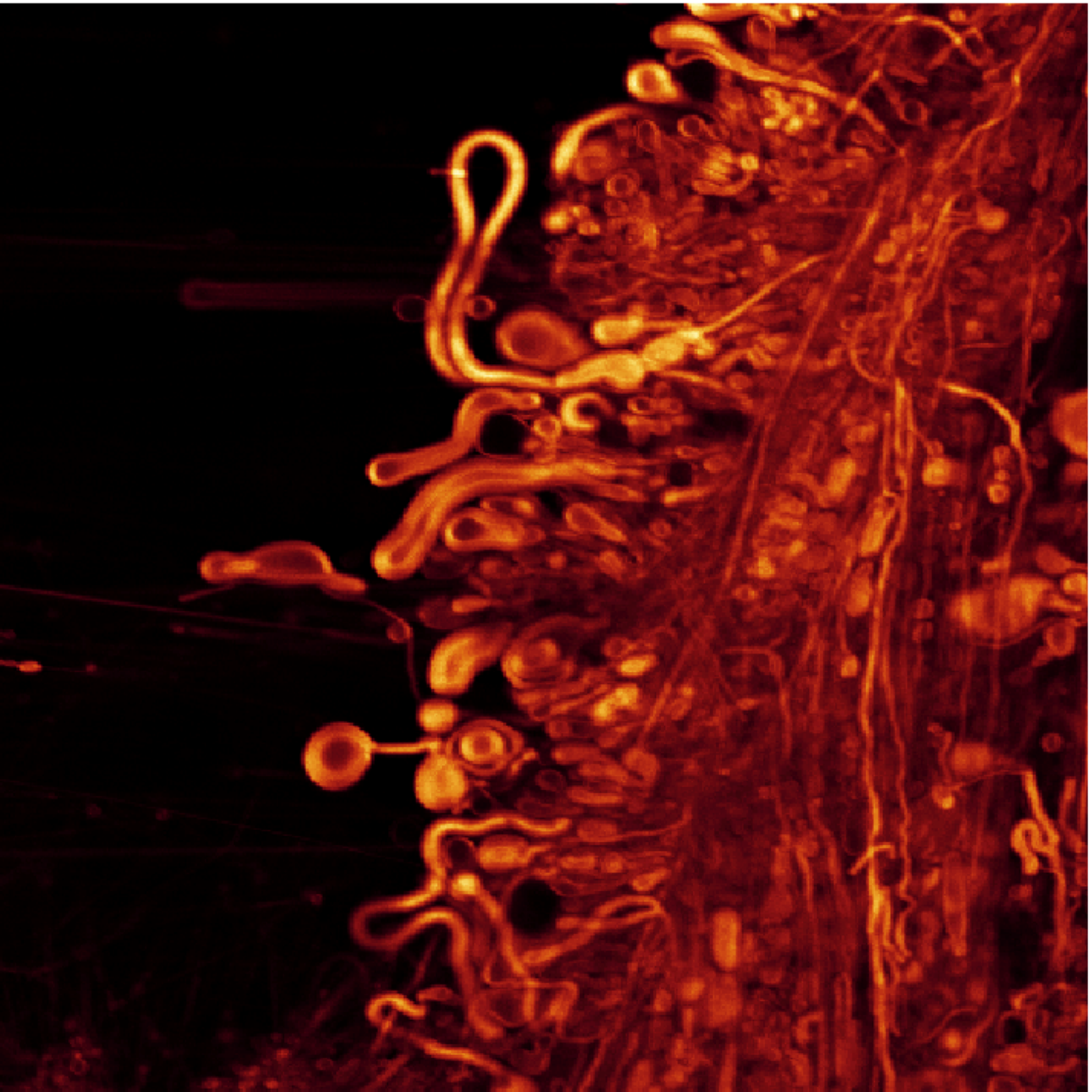}
\caption{\label{fig.0} Different size and morphology of multilamellar tubules (MLTs) dispersed in excess water. The cross-sectional view ($279 \ \mu$m on each side) is obtained by fluorescence confocal polarized microscopy (FCPM) with incident laser polarization oriented along the vertical axis of the image.}
\label{figure7}
\end{figure}

\section*{Conclusions}
Stability of MLTs implies that they are in hydrostatic equilibrium and stable as elastic objects that have compression and bending elasticity. In this regime, the model is universal and independent of the underlying growth mechanisms of MLTs. The MLTs did not exhibit any significant motion on the typical time scales of the experiments in the closed sample cell. Furthermore, in the stability regime, MLTs are treated as aggregates of lipid bilayers with an equilibrium spacing of $D_{\rm{eq}}$ between adjacent bilayers, which is achieved by balancing the intermolecular forces within the stack. In the presence of defects, growth is driven by the hydration gradient.\cite{TB2021} The size and morphology of MLTs in the experiments cannot be controlled, as shown in Fig. \ref{figure7}. However, if the lamellar stacks are spin coated, then MLTs do not form.\cite{BHM}  Multilateral tubule growth, also known as myelin figures (MF) \cite{CF}, during lipid hydration is explored using surfactants such as Aerosol-OT, triethylene glycol monododecyl ether \cite{JB2003,MK1987,HZW,SSS,SK85,ZN,Z2009,MB2000,SE2010}, from vesicular dispersions of lung surfactant extract \cite{ES2018}, lipid vesicles with hydrophilic polymers \cite{TS2003,TS2001} and aqueous multicomponent polymer solutions \cite{GJC1980}. In living systems, an important example is the presence of MF in the pulmonary lining, an extracellular lipid-protein coat on the alveolar surface, the presence of which helps reduce surface tension when the lungs are deflated \cite{SRJ, VLMG, WER}. 
The model is simple and might be of interest to the field in understanding tubule stability in a variety of biological processes/applications. The results are shown in Fig.~\ref{fig.4} and show that even with solvent pressure differences of 0.1 atm, the density profile is not significantly altered, thus determining that the stability is due to the trapped solvent. I also believe that the results are important enough for the lamellar stability model to be considered for publication, with implications beyond the MLT models described, especially when trying to model other membrane systems.

\section*{Acknowledgment}
TB acknowledges a Ph.D. thesis research fellowship from the Raman Research Institute (RRI) , Bangalore, India. TB acknowledges Prof. Yashodhan Hatwalne (RRI) , Dr. A. Chaudhuri (IISERM), and Prof. J. H. Ipsen (SDU)  for useful discussions.

\footnotesize{
\bibliography{main.bib} 
\bibliographystyle{rsc} 
}

\end{document}